\documentclass[10pt]{article}
\usepackage{graphicx}
\makeatletter
\def\thesection{\arabic{section}}
\@addtoreset{equation}{section}

\def\appendix{\par
\setcounter{section}{0}
\setcounter{subsection}{0}
\def\thesection{\Alph{section}}}
\makeatother
\DeclareSymbolFont{boldletters}{OML}{cmm} {b}{it}
\DeclareSymbolFontAlphabet{\mathbit}{boldletters}
\DeclareMathSymbol{\alpha}{\mathalpha}{letters}{"0B}
\DeclareMathSymbol{\beta}{\mathalpha}{letters}{"0C}
\DeclareMathSymbol{\gamma}{\mathalpha}{letters}{"0D}
\DeclareMathSymbol{\delta}{\mathalpha}{letters}{"0E}
\DeclareMathSymbol{\epsilon}{\mathalpha}{letters}{"0F}
\DeclareMathSymbol{\zeta}{\mathalpha}{letters}{"10}
\DeclareMathSymbol{\eta}{\mathalpha}{letters}{"11}
\DeclareMathSymbol{\theta}{\mathalpha}{letters}{"12}
\DeclareMathSymbol{\iota}{\mathalpha}{letters}{"13}
\DeclareMathSymbol{\kappa}{\mathalpha}{letters}{"14}
\DeclareMathSymbol{\lambda}{\mathalpha}{letters}{"15}
\DeclareMathSymbol{\mu}{\mathalpha}{letters}{"16}
\DeclareMathSymbol{\nu}{\mathalpha}{letters}{"17}
\DeclareMathSymbol{\xi}{\mathalpha}{letters}{"18}
\DeclareMathSymbol{\pi}{\mathalpha}{letters}{"19}
\DeclareMathSymbol{\rho}{\mathalpha}{letters}{"1A}
\DeclareMathSymbol{\sigma}{\mathalpha}{letters}{"1B}
\DeclareMathSymbol{\tau}{\mathalpha}{letters}{"1C}
\DeclareMathSymbol{\upsilon}{\mathalpha}{letters}{"1D}
\DeclareMathSymbol{\phi}{\mathalpha}{letters}{"1E}
\DeclareMathSymbol{\chi}{\mathalpha}{letters}{"1F}
\DeclareMathSymbol{\psi}{\mathalpha}{letters}{"20}
\DeclareMathSymbol{\omega}{\mathalpha}{letters}{"21}
\DeclareMathSymbol{\varepsilon}{\mathalpha}{letters}{"22}
\DeclareMathSymbol{\vartheta}{\mathalpha}{letters}{"23}
\DeclareMathSymbol{\varpi}{\mathalpha}{letters}{"24}
\DeclareMathSymbol{\varrho}{\mathalpha}{letters}{"25}
\DeclareMathSymbol{\varsigma}{\mathalpha}{letters}{"26}
\DeclareMathSymbol{\varphi}{\mathalpha}{letters}{"27}
\DeclareMathSymbol{\Gamma}{\mathalpha}{letters}{"00}
\DeclareMathSymbol{\Delta}{\mathalpha}{letters}{"01}
\DeclareMathSymbol{\Theta}{\mathalpha}{letters}{"02}
\DeclareMathSymbol{\Lambda}{\mathalpha}{letters}{"03}
\DeclareMathSymbol{\Xi}{\mathalpha}{letters}{"04}
\DeclareMathSymbol{\Pi}{\mathalpha}{letters}{"05}
\DeclareMathSymbol{\Sigma}{\mathalpha}{letters}{"06}
\DeclareMathSymbol{\Upsilon}{\mathalpha}{letters}{"07}
\DeclareMathSymbol{\Phi}{\mathalpha}{letters}{"08}
\DeclareMathSymbol{\Psi}{\mathalpha}{letters}{"09}
\DeclareMathSymbol{\Omega}{\mathalpha}{letters}{"0A}

\newcommand{\mbit}[1]{{\mathbit#1}}
\newcommand{\Nabla}{{\mbox{\boldmath$\nabla$}}}
\newcommand{\bra}[1]{\left\langle{#1}\right\vert}
\newcommand{\ket}[1]{\left\vert{#1}\right\rangle}
\newcommand{\braket}[2]{\left\langle{#1}\vert{#2}\right\rangle}

\newcommand{\dsl}[2][0pt]{#2{\mbox{\hspace{-8pt}\hspace{#1}$\not$}
\hspace{8pt}\hspace{-#1}}} 
\newcommand{\dslp}{\dsl[1.5pt]{p}}
\newcommand{\dslk}{\dsl{k}}
\newcommand{\dslpart}{\dsl{\partial}}

\newcommand{\ddd}[2]{\mathop{\partial_{{#1}{}}^{#2}}}
\newcommand{\der}[2]{\ddd{#1}{#2}\!}

\newcommand{\DDD}[2]{\mathop{D_{{#1}{}}^{#2}}}

\newcommand{\lrDer}[2]{\DDD{#1}{#2}^{\leftrightarrow}\!}
\newcommand{\lrDerb}{\mathop{\dsl{D}{}}^{\leftrightarrow}\!}
\newcommand{\dbox}{\,\framebox(7,7)[t]{}\,}
\begin{document}    
\title{Gauge Independence in terms of 
the Functional Integral\thanks{KYUSHU-HET-36}}
\author{Taro~KASHIWA\thanks{taro1scp@mbox.nc.kyushu-u.ac.jp}~~and 
Naoki~TANIMURA\thanks{tnmr1scp@mbox.nc.kyushu-u.ac.jp}  \\   
Department of Physics, Kyushu University \\ Fukuoka 812-81, 
JAPAN 
}
\maketitle
\begin{abstract}
\noindent Among various approaches in proving gauge independence, models 
containing an explicit gauge dependence are convenient. The well-known 
example is the gauge
parameter in the covariant gauge fixing which is of course most 
suitable for the perturbation theory but a negative metric prevents us 
from imaging a dynamical picture. Noncovariant gauge
such as the Coulomb gauge is on the contrary used for many physical 
situations. Therefore it is desirable to include both cases. More than ten years ago, 
Steinmann introduced a function (distribution) which can play this 
role in his attempt on discussing quantum electrodynamics (QED) in terms
of the gauge invariant fields solely. The method is, however, broken 
down 
in the covariant case: the invariant 
operators are ill-defined because of $1/ p^2$ singularity in the Minkowski 
space. In this paper, we apply his function to the path 
integral: utilizing the arbitrariness of the function we first 
restrict it to be able to have a well-defined operator, and then a 
Hamiltonian with which we can
build up the (Euclidean) path integral formula. Although the 
formula is far from covariant, a full covariant expression is recovered
by reviving the components which have been discarded under the 
construction of the Hamiltonian. There is no pathological defects 
contrary to the operator formalism. 
With the aid of the path integral formula, 
the gauge independence of the free energy as well as the S-matrix is proved. 
Moreover the reason is clarified 
why it is so simple and straightforward to argue gauge
transformations in the path integral. Discussions on the quark 
confinement is also presented. 
\end{abstract}

\newpage

\section{Introduction} 
Gauge transformations in quantum mechanics reads as the unitary 
transformation so that proving gauge invariance is nothing 
but proving the unitary equivalence between two theories\cite{PTF}. 
However
in quantum field theories, it is well known that canonical commutation 
relations demand to fix the gauge, that is, each gauge has its own 
Hilbert space. Consequently in order to assert that the result is gauge 
independent, we usually compare results which have been obtained by 
different gauges\cite{BB}. (Some approaches, however, treat 
gauge transformation itself even in quantum field theories\cite{KFC}.)  
Therefore, models with an explicit gauge dependence may be suitable;
for instance $\alpha$ in the 
Nakanishi-Lautrup formalism\cite{NL}, 
\begin{equation}
{\cal L} = - {1 \over 4} F_{\mu
\nu} F^{\mu \nu} - A^\mu \partial_\mu B + {\alpha \over 2} {B}^2 \ .	
	\label{1a}
\end{equation}
(Throughout the paper, 
repeated indices are implied summation unless otherwise stated.) 
Covariance is indispensable 
in perturbation theories but requires the negative metric which makes 
it hard to imagine dynamics. On the contrary, noncovariant, such as 
the Coulomb or the axial gauge can be formulated in an ordinary Hilbert 
space with a positive metric; which is more useful in 
physical situations\cite{BNS}.  Therefore it is desirable for a model to include 
both cases. 

More than ten years ago, Steinmann\cite{ST} discussed quantum 
electrodynamics (QED) perturbatively in terms of gauge invariant 
fields which are defined by the operators, $\psi$, $\overline\psi$, 
and $A^\mu$ in the Gupta-Bleuler formalism:
\begin{eqnarray}
\Psi(x)\!&\!\equiv\!&\!\exp\left[-ie\int\!\!d^{4}y
\phi^\mu(x-y)A_\mu(y)\right]\psi(x)  \ ,   \nonumber \\
\noalign{\vskip -1ex} 
\!\!& \!\! & \!\!     \label{psi}   \\
\noalign{\vskip -1ex}  
\overline\Psi(x)\!&\!\equiv\!&\!\Psi^{\dag}\gamma_0 \ , \nonumber 
\end{eqnarray}
with $\phi^\mu(x)$ being real function (distribution strictly 
speaking) satisfying
\begin{equation}
\partial_\mu\phi^\mu(x)=\delta^4(x) \ . \label{phi}
\end{equation}
Therefore by noting
\begin{equation}
F^{\mu\nu} \equiv \partial^\mu A^\nu -\partial^\nu A^\mu , \ 	 
\label{a4}
\end{equation}
 all the equations of motion become gauge invariant:
\begin{eqnarray}
   \partial^\nu F_{\mu\nu}(x) \!\! & = \!\! & ej_\mu(x) \ ,
 \\
(i\dslpart-m)\Psi(x)	 \!\!  & =\!\! & e\gamma^\mu\int\!\!d^4y
\phi^\nu(x-y)F_{\mu\nu}(y)
\Psi(x)  \ ,  
\end{eqnarray}
It was then shown that all the Wightman functions can be calculated perturbatively. 
Contrary to his intention that the result must be gauge invariant 
but that is of course $\phi_{\mu}$-dependent. Indeed we can
regard his method as the $\phi$-gauge fixing\cite{KaTa}; which
can be seen by noting 
the following quantity,
\begin{equation}
\bra{0}T^*{A}^{\lambda_{1}}(x_{1};\phi)\cdots {A}^{\lambda_{n}}(x_{n};\phi)
        \Psi(y_{1})\cdots\Psi(y_{m})
        \overline\Psi(z_{1})\cdots\overline\Psi(z_{m})
        \ket{0}  \ ,   \label{tf}
\end{equation}
with $T^*$ designating the covariant $T^*$-product, 
where $A^\mu(x;\phi)$ is physical, that is, gauge invariant photon
field, given by 
\begin{equation}
A_{\mu}(x;\phi)\equiv-\int\!\!d^4y\phi^\nu(x-y)F_{\mu\nu}(y) \ .
\label{aphi1}
\end{equation}
By a perturbative calculation we can see that the original photon 
propagator,
 \begin{equation}
D^{\mu\nu}(q)  \equiv{-i \over q^2 }
\left( g_{\mu \nu } - {q_{\mu}q_{\nu} \over  q^2 } 
	\right) 
 \end{equation}
must be replaced to the $\phi$-dependent one,
 \begin{equation}
     D^{\mu\nu}(q; \phi) \equiv  {-i\over 
     q^{2}}\Bigl\{g^{\mu\nu}+iq^{\mu}
        \widetilde\phi^{\nu}(q)
        - i{\widetilde\phi^{\mu*}} (q)q^{\nu}
        + q^{\mu}q^{\nu}\vert\widetilde\phi(q)\vert^{2}\Bigr\} \ ,  
        \label{propag}
\end{equation}
with $\widetilde\phi_\mu(q)$ being the Fourier transform of
$\phi_\mu(x)$, in the ordinary n-point function:
\begin{equation}
\bra{0}TA^{\lambda_{1}}(x_{1})\cdots A^{\lambda_{n}}(x_{n})
        \psi(y_{1})\cdots\psi(y_{m})
        \overline\psi(z_{1})\cdots\overline\psi(z_{m})
        \ket{0} \ . 
\end{equation}

Since $\phi_{\mu}$ is an arbitrary function, the choices,  
\begin{equation}
	\phi^\mu(x)
	= \Bigl( 0,  {\Nabla \over \Nabla^{2}}\delta(x^{0}) \Bigr)\equiv \Bigl( 0,  {\mbit{x}\over 4 \pi |\mbit{x}|^{3}} \delta(x^{0}) 
	\Bigr)  \ ,
	\label{coulomb}
\end{equation}
\begin{equation}
	\phi^\mu(x)
	= \Bigl( 0,  0,  0, \delta(x^{0}) \delta(x^{1}) \delta(x^{2}) 
	\theta(x^{3}) \Bigr)  \ , 
	\label{axial}
\end{equation}
and 
\begin{equation}
  \phi_\mu(x)   =   {\partial_{\mu} \over  \dbox} \  ,  \qquad 
  \Bigl(	\widetilde\phi_\mu(q)  =   { iq_{\mu}  \over   q^{2} } \ , 
  \Bigr)
\end{equation}
give us the Coulomb, the axial, and the covariant Landau gauge, 
respectively.  However, there is a problem in dividing $q^{2}$; 
which brings $\Psi(x)$ and 
$\overline\Psi(x)$ to ill-defined operators. If we adopt $1/(q^{2}-i\epsilon)$ 
prescription to avoid the singularity, the reality of $\phi_{\mu}(x)$ 
is lost. 
If then we use $1/(q^{2}+ i\epsilon q_{0})$, or $1/(q^{2}- i\epsilon q_{0})$, 
or the principal value prescription for preserving the reality, time 
reversal invariance or causality is broken down.  Actually the support 
of $\phi_{\mu}(x)$ must be spacelike, since the timelike support causes 
the difficulty in the ordering between $\psi$ and $\overline\psi$ and 
$A_{\mu}$. Therefore as far as the operator formalism is concerned, 
covariance 
is superficial in the Steinmann's approach. 

Contrary to the operator formalism, it is known that the path 
integral formalism can handle gauge invariance more 
efficiently. Furthermore, the Euclidean path integral 
expression, when an imaginary time goes to infinity\cite{AL},
contains all the information of the ground states of the 
theory. 
In this study, we start 
with the spacelike $\phi_{\mu}$ so as to throw away the redundant 
variables and perform canonical quantization without any problem. 
Then the Euclidean path integral expression is built 
by the trace formula of an imaginary time evolution operator; which 
is, of course, far from covariant. At the final stage, the 
redundant variables are revived by means of insertions of some 
identities into the path integral\cite{KS}. In the formula, we can take 
any choice of $\phi_{\mu}$ by means of the Faddeev-Popov trick.
In this way, all the above difficulties
can be avoided. These are the contents of \S2. In \S3, the proof for gauge 
independence of the free energy and the 
S-matrix is presented. In the next \S4, we 
clarify the reason for ability of discussing the gauge invariance 
more straightforwardly in 
the functional method; since contrary to many discussions on 
gauge transformations using the path integral, there seems to have been no 
close examination on justification. The final \S 5 is devoted to discussions.

\section{Euclidean Path Integral Expression for an Arbitrary Gauge Fixing}
In this section, we construct the Euclidean path integral 
expression by first reducing the gauge degree of freedom with the aid 
of $\phi_{\mu}$, then applying canonical quantization to this highly 
noncovariant system. By utilizing the functional 
representation\footnote{The
 path integral formula by 
means of the holonomic representation\cite{FS} always suffers from
the nonlocal bilinear term in relativistic cases due to anti-particle 
contributions. Therefore it is unsuitable to handle with gauge 
transformations.}, 
the (noncovariant) path integral expression is obtained, which finally 
comes back to a covariant form by reviving the redundant degrees in 
terms of
insertions of some identities. 

Assume that the support of $\phi_\mu(x)$ are spacelike:
\begin{equation}
\phi_\mu(x)=(0,f_i({\mbit x})\delta(x_0)) \ . \label{phi1}
\end{equation}
In the momentum space,
\begin{equation}
\widetilde\phi_\mu(p)=(0,\widetilde f_i({\mbit p}))  \ ,  \label{phi2}
\end{equation}
with
\begin{equation}
\phi_\mu(x)=\int\!\!{d^4p\over(2\pi)^4}{\rm e}^{ipx}
\widetilde\phi_\mu(p) \ , 
\end{equation}
then the reality of $\phi_\mu(x)$ and the relation (\ref{phi}) turn 
out to be
\begin{equation}
\widetilde\phi_\mu^*(p)=\widetilde\phi_\mu(-p) \ ,\hspace{20pt}
p^\mu\widetilde\phi_\mu(p)=i \ ,
\end{equation}
respectively.
With the help of $\phi_\mu$'s, we can extract the gauge degree of 
freedom from $A^\mu$ such that
\begin{equation}
A^\mu(x)=A^\mu(x;\phi)+\partial^\mu\omega(x) \ ,\label{dec}
\end{equation}
where 
\begin{eqnarray}
A^\mu(x;\phi)\!&\!\equiv\!&\!\int\!\!d^4y\{\delta^\mu_\nu\delta^4(x-y)
-\partial^\mu_x\phi_\nu(x-y)\}A^\nu(y) \ , \label{aphi2} \\
\omega(x)\!&\!\equiv\!&\!\int\!\!d^4y\phi^\mu(x-y)A_\mu(y) \ . 
\label{omega}
\end{eqnarray}
The relation (\ref{aphi2}) is nothing but (\ref{aphi1}) with the use of 
(\ref{a4}) as well as (\ref{phi}) . In order to discuss path integral 
the Schr\"{o}dinger picture is employed 
so that the time argument of the fields is omitted, thus $A^i({\mbit x})$ is used 
for the time being. Therefore the relations, (\ref{dec}) to (\ref{omega}), read as
\begin{eqnarray}
A^i({\mbit x}) \!&\! = \!&\! A^i({\mbit x};\phi)+\Nabla^i\omega({\mbit x}) \ ,  \\ 
A^i({\mbit x};\phi)\!&\! =\!&\!\int\!\!d^3y\{\delta^{ij}
\delta^3({\mbit x}-{\mbit y})-\Nabla^i_x 
f^j({\mbit x}-{\mbit y})\}A^j({\mbit y}) \ , \label{b9} \\
\omega({\mbit x})\!&\! =\!&\!\int\!\!d^3yf^j({\mbit x}-{\mbit y})
A^j({\mbit y})  \  . \label{b10}
\end{eqnarray}
Since a genuine 
physical component must have two components, we must select them out 
from three $A^i({\mbit x})$'s: to this end consider the norm of functional
space of $A^i({\mbit x})$: 
\begin{eqnarray}
\int\!\!d^3x \ \delta A^i({\mbit x}) \ \delta A^i({\mbit x})
\!&\!=\!&\!\int\!\!d^3x\{\delta A^i({\mbit x};\phi)\delta 
A^i({\mbit x};\phi)
-\delta\omega({\mbit x})\Nabla^i\delta A^i({\mbit x};\phi)\nonumber\\
\!&\!\!&\!\hspace{30pt} + \ \delta A^i({\mbit x} ;\phi)
\Nabla^i\delta\omega({\mbit x})
-\delta\omega({\mbit x})\Nabla^2\delta\omega({\mbit x})\} \ , \label{norm}
\end{eqnarray}
whose $\delta A^i(x;\phi) \ \delta A^i(x;\phi)$ part reads
\begin{equation}
\int\!\!d^3x \ \delta A^i({\mbit x};\phi) \ \delta A^i({\mbit x};\phi)
=\int\!\!{d^3p\over(2\pi)^3}\delta\widetilde A^{j*}({\mbit p})
M^{jk}({\mbit p})\delta\widetilde A^k({\mbit p})  \ , 
\end{equation}
with $\widetilde A^k({\mbit p})$ satisfying 
$\widetilde A^k(-{\mbit p})=\widetilde A^{k*}({\mbit p})$. Here 
\begin{equation}
M^{jk}({\mbit p})\equiv\delta^{jk}+i\widetilde f^{j*}({\mbit p})p^k
-ip^j\widetilde f^k({\mbit p})
+{\mbit p}^2\vert\widetilde{\mbit f}({\mbit p})\vert^2 \ 
\end{equation}
which can be diagonalized as\cite{Tab}
\begin{equation}
n_{(\alpha)}^j({\mbit p})M^{jk}({\mbit p})n_{(\beta)}^{k*}({\mbit p})=\left(
\begin{array}{ccc}
1&&\\&{\mbit p}^2\vert\widetilde{\mbit f}({\mbit p})\vert^2&\\&&0
\end{array}
\right)_{\alpha\beta}  \ ,    \label{b14}
\end{equation}
where 
\begin{eqnarray}
n_{(1)}^k({\mbit p})\!&\!\equiv\!&\!\epsilon^{klm}
n_{(2)}^l({\mbit p})n_{(3)}^m({\mbit p}) \  ,\nonumber\\
n_{(2)}^k({\mbit p})\!&\!\equiv\!&\!\{ip^k+{\mbit p}^2\widetilde 
f^k({\mbit p})\}
\Big/\sqrt{{\mbit p}^2({\mbit p}^2\vert\widetilde{\mbit f}({\mbit 
p})\vert^2-1)} \ ,
\label{ns}    \\
n_{(3)}^k({\mbit p})\!&\!\equiv\!&\!ip^k/\vert{\mbit p}\vert \ ,\nonumber
\end{eqnarray} 
are the orthonormal base obeying 
\begin{equation}
\sum_{\alpha=1}^3n_{(\alpha)}^{j*}({\mbit p})n_{(\alpha)}^k({\mbit p})
=\delta^{jk}, \hspace{20pt}n_{(\alpha)}^{k*}({\mbit p})
=n_{(\alpha)}^k(-{\mbit p}) \ . 
\end{equation}
In view of (\ref{b14}) we can take the desired physical 
components as
$\widetilde A^{(1)}({\mbit p})$ and $\widetilde A^{(2)}({\mbit p})$ 
where
\begin{equation}
\widetilde A^{(\alpha)}({\mbit p}) \equiv n_{(\alpha)}^k({\mbit p})
  \widetilde A^k({\mbit p}) ;  \hspace{15ex}  ( \alpha=1,2)  \ .	
\end{equation}
Therefore according to (\ref{b9}) $A^i({\mbit x};\phi)$ is now expressed solely
 by the physical components:
 \begin{eqnarray}
A^i({\mbit x};\phi)\!&\!=\!&\!\int\!\!{d^3p\over(2\pi)^3}
{\rm e}^{i{\mbit p\cdot \mbit x}}
\{n_{(1)}^{i*}({\mbit p})\widetilde A^{(1)}({\mbit p})
+[\delta^{ij}-ip^i\widetilde f^j({\mbit p})]n_{(2)}^{j*}({\mbit p})
\widetilde A^{(2)}({\mbit p})\}     \nonumber\\
\!&\!=\!&\!n_{(1)}^{i*}(-i\Nabla)A^{(1)}({\mbit x})
+[\delta^{ij}-\Nabla^i\widetilde f^j(-i\Nabla)]
n_{(2)}^{j*}(-i\Nabla)A^{(2)}({\mbit x}) \ ,  \label{aphi3}
\end{eqnarray}
where as usual
\begin{equation}
	 A^{(\alpha)}({\mbit x})  \equiv  \int\!\!{d^3p\over(2\pi)^3}
{\rm e}^{i{\mbit p\cdot \mbit  x}}\widetilde A^{(\alpha)}({\mbit 
p}) \ ; 
\hspace{10ex}  (\alpha=1,2)  \ .     
\end{equation}
Finally the norm (\ref{norm}) is given by
\begin{eqnarray}
\lefteqn{
\int\!\!d^3x\delta A^i({\mbit x})\delta A^i({\mbit x})
}\nonumber\\
\!&\!=\!&\!\int\!\!d^3x
\left(\begin{array}{ccc}
\!\!\delta A^{(1)}\!\!&\!\!\delta A^{(2)}\!\!&\!\!\delta\omega\end{array}
\!\!\right)
\!\!\left(\!\begin{array}{ccc}1&0&0\\
0&-\Nabla^2\vert\widetilde{\mbit f}(-i\Nabla)\vert^2
 &-\sqrt{
\hspace{15pt}\bullet \hspace{15pt}}\\
0&-\sqrt{
\hspace{15pt}\bullet \hspace{15pt}}
 &-\Nabla^2
\end{array}\!\!\right)\!\!
\left(\!
\!\begin{array}{c}\delta A^{(1)}\\\delta A^{(2)}\\\delta\omega
\end{array}\!\!\right) \ ,   \label{norm2}
\end{eqnarray}
where 
\begin{equation}
	\sqrt{\hspace{15pt}\bullet \hspace{15pt}}\equiv
	\sqrt{\Nabla^2(\Nabla^2\vert\widetilde{\mbit 
	f}(-i\Nabla)\vert^2+1)} \ .
\end{equation}

The action, with the source term $J^{\mu}$, reads 
\begin{eqnarray}
S\!&\!=\!&\!\int\!\!d^{4}x\,\left\{-{1\over4}
F^{\mu\nu}F_{\mu\nu}+J^{\mu}A_{\nu}\right\}\nonumber\\
\!&\!=\!&\!\int\!\!d^{4}x\,\Biggl\{{1\over2}\sum_{i,j=1}^{3}
A^{i}(x;\phi)(\delta^{ij}\Nabla^{2}-\Nabla^{i}\Nabla^{j})A^{j}(x;\phi)
+{1\over2}\stackrel{\mbox{\Large.}}{\mbit A}^{2}\!\!\!(x;\phi)
-A^{0}(x;\phi) \Nabla\cdot\!\!\stackrel{\mbox{\Large.}}{\mbit A}\!\!(x;\phi)
\nonumber\\
\!&\!\!&\! +\hspace{0pt}{1\over2}(\Nabla A^{0}(x;\phi))^{2}
+J_{0}(x) A^{0}(x;\phi)
-{\mbit J}(x) \!\cdot\!{\mbit A}(x;\phi)\Biggr\}  \ ,  \label{b21}
\end{eqnarray}
which is further rewritten, with the help of (\ref{aphi3}) as well as 
(\ref{ns}), as
\begin{eqnarray}
\!\!\!\!\!\!\!\!S\!&\!\!\!=\!\!\!&\!\!\!\int\!\!d^{4}x\,\Biggl\{{1\over2}\sum_{\alpha=1}^{2}
A^{(\alpha)}(x)\Nabla^{2}A^{(\alpha)}(x)
+{1\over2}\Bigl(\stackrel{\mbox{\Large.}}{A}^{(1)}\!\!\!(x)\Bigr)^{2}
-{1\over2}{\stackrel{\mbox{\Large.}}{A}^{(2)}}\!\!\!(x)\!\Nabla^{2}
\vert\widetilde{\mbit f}(-i\Nabla)\vert^{2}{\stackrel{\mbox{\Large.}}{A}^{(2)}}
\!\!\!(x)
\nonumber\\
\!\!\!\!\!\!\!\!\!\!\!\!\!&\!\!&\! \!\!\!\!-  A^{0}(x;\phi)   
\sqrt{\Nabla^2(\Nabla^2\vert\widetilde{\mbit f}(-i\Nabla)\vert^2+1)}
{\stackrel{\mbox{\Large.}}{A}^{(2)}}\!\!\!(x) 
+{1\over2}(\Nabla A^{0}(x;\phi))^{2}+J_{0}(x)A^{0}(x;\phi)
-{\mbit J}(x) \!\cdot\! \Bigl({\mbit A}(x;\phi)\Bigr)\Biggr\} \ ,  \label{b22}
\end{eqnarray}
whose last term, $\bigl( {\mbit A}(x;\phi) \bigr)$, implies 
that ${\mbit A}(x;\phi)$ has been given by $A^{(1)}$ and $A^{(2)}$ by 
the relation (\ref{aphi3}). 
The system still remains in a constrained system because $A^{0}(x;\phi)$ 
is not a dynamical variable. Solving the constraint, we obtain the 
Hamiltonian:
\begin{eqnarray}
H(t) \!&\!=\!&\!\int\!\!d^{3}x\,\Biggl\{{1\over2}\sum_{\alpha=1}^{2}
\left[\Bigl(\Pi^{(\alpha)}({\mbit x})\Bigr)^{2}+\Bigl(\Nabla A^{(\alpha)}({\mbit x})
\Bigr)^{2}\right]
\nonumber\\
\!&\!\!&\!+J_{0}(x)
{\sqrt{\Nabla^2(\Nabla^2\vert\widetilde{\mbit f}(-i\Nabla)\vert^2+1)}
\over\Nabla^{2}}\Pi^{(2)}({\mbit x})
+{1\over2}J_{0}(x) \vert\widetilde{\mbit f}(-i\Nabla)\vert^{2}J_{0}(x)
+{\mbit J}(x) \!\cdot\!({\mbit A}({\mbit x};\phi))\Biggr\} \ ,
\end{eqnarray}
where $x \equiv (t, {\mbit x})$ and $\Pi^{(\alpha)}$ is the canonical conjugate 
momentum of $A^{(\alpha)}$. Note that the explicit time dependence lies in 
the sources so as to indicate the Hamiltonian.

Now quantization can be carried out by means of the canonical commutation 
relations:
\begin{equation}
[\hat A^{(\alpha)}({\mbit x}),\hat\Pi^{(\beta)}({\mbit y})]
=i\delta^{\alpha\beta}\delta^{3}({\mbit x-\mbit y}) \ ,\hspace{10pt}
[\hat A^{(\alpha)}({\mbit x}),\hat A^{(\beta)}({\mbit y})]
=[\hat\Pi^{(\alpha)}({\mbit x}),\hat\Pi^{(\beta)}({\mbit y})]
=0 \ ,
\end{equation}
where caret denotes an operator. The partition function
 in our interest is defined by 
\begin{equation}
Z_{T}[J]\equiv\lim_{N\rightarrow\infty}{\rm Tr}
\left[({\bf I}-\Delta\tau\hat H_{N})({\bf I}-\Delta\tau\hat H_{N-1})
\cdot\cdot\cdot({\bf I}-\Delta\tau\hat H_{1})\right] \ ,
\end{equation}
where $\Delta\tau\equiv T/N$, $H_{j}\equiv H(j\Delta\tau)$, and 
the source, $J_{\mu}(x)$, has been assumed to be analytically 
continuable. Here Tr can be taken for any complete set, but a functional 
representation,
\begin{equation}
	\hat A^{(\alpha)}({\mbit x})\ket{ \{A\} }
	=A^{(\alpha)}({\mbit x})\ket{ \{A\} },\hspace{20pt}
	\hat\Pi^{(\alpha)}({\mbit x})\ket{ \{ \Pi \}}
	=\Pi^{(\alpha)}({\mbit x})\ket{ \{ \Pi \} } \ ,
	\label{b26}
\end{equation}
with completeness, 
\begin{equation}
\int\!\!{\cal D}A^{(\alpha)}\ket{\{A\}}\bra{\{A\}}={\bf I} \ ,\hspace{20pt}
\int\!\!{\cal D}\Pi^{(\alpha)}\ket{\{ \Pi \}}\bra{\{ \Pi \}}={\bf 
I} \ ,  \label{b27}
\end{equation}
should be chosen so as to obtain the Euclidean path integral representation\cite{KS}:
\begin{eqnarray}
Z_{T}[J]\!&\!=\!&\!\lim_{N\rightarrow\infty}{\cal N}^{2N}\int\!\!
\prod_{k=1}^{N}{\cal D}A_{k}^{(\alpha)}{\cal D}\Pi_{k}^{(\alpha)}
\nonumber\\
\!&\!\!&\!\times\exp\left[\Delta\tau\sum_{k=1}^{N}
\left\{\int\!\!d^{3}x\,i\sum_{\alpha=1}^{2}\Pi_{k}^{(\alpha)}
({\mbit x})
{A_{k}^{(\alpha)}({\mbit x})-A_{k-1}^{(\alpha)}({\mbit x})\over\Delta\tau}
-H_{k}(\Pi_{k}^{(\alpha)},A_{k}^{(\alpha)})\right\}\right] \ , 
\end{eqnarray}
where $\cal N$ is the normalization factor defined in 
\begin{equation}
\braket{\{ \Pi \}}{\{A \} } \equiv {\cal N}\exp\left[-i\int\!\!d^{3}x
\sum_{\alpha=1}^{2}\Pi^{(\alpha)}({\mbit x})A^{(\alpha)}({\mbit x})
\right] \  .   \label{pia}
\end{equation}
More explicitly
\begin{equation}
{\cal N}=\prod_{\mbit x}{1\over2\pi} \ .\label{N}
\end{equation}
Due to the trace, the boundary condition is periodic $A_{0}^{(\alpha)}
({\mbit x})=A_{N}^{(\alpha)}({\mbit x})$.
(This is a formal expression and is ill-defined actually. To make 
$Z_{T}[J]$ well-defined, it is also necessary to discretize the 
space part and to put the system in a box. For gauge theories, 
there still remains the 
difficulty of the infrared divergence which can, however, be avoided 
by making further 
considerations\cite{KS}. Hereafter we confine ourselves in a continuous 
representation for notational simplicity.)

Therefore we write
\begin{eqnarray}
Z_{T}[J]\!&\!=\!&\!\int\!\!{\cal D}A^{(\alpha)}
{\cal D}\Pi^{(\alpha)}\exp\left[\int\!\!d^{4}x_{_{E}}
\left\{ \, i \sum_{\alpha=1}^{2}\Pi^{(\alpha)}
\stackrel{\mbox{\Large.}}{A}^{(\alpha)}- H(\Pi, A)\right\}\right]
\nonumber\\
\!&\!=\!&\!\int\!\!{\cal D}A^{(\alpha)}
{\cal D}\Pi^{(\alpha)}\exp\Biggl[\int\!\!d^{4}x_{_{E}}
\Biggl\{ i \sum_{\alpha=1}^{2}
\Pi^{(\alpha)}(\tau,{\mbit x}) 
\stackrel{\mbox{\Large.}}{A}^{(\alpha)}\!\!\!(\tau,{\mbit x})  
\nonumber \\
\!&\! \! & \! -{1\over2}\sum_{\alpha=1}^{2}\left[ \Bigl( \Pi^{(\alpha)}
(\tau,{\mbit 
x}) \Bigr)^{2}+\Bigl( A^{(\alpha)}(\tau,{\mbit x})\Bigr)^{2}
\right]  + iJ_{4}(\tau,{\mbit x}){\sqrt{\Nabla^2(\Nabla^2\vert\widetilde{\mbit f}
(-i\Nabla)\vert^2+1)}\over\Nabla^{2}} \Pi^{(2)}(\tau,{\mbit 
x})   \nonumber\\
\!&\!\!&\! 
+{1\over2}J_{4}(\tau,{\mbit x})\vert\widetilde{\mbit f}(-i\Nabla)\vert^{2}
J_{4}(\tau,{\mbit x})
-{\mbit J}(\tau,{\mbit x})\!\cdot\!({\mbit A}
(\tau,{\mbit x};\phi))\Biggr\}\Biggr],
\end{eqnarray}
where 
\begin{eqnarray}
	\int\!\!d^{4}x_{_{E}} \!\! & \equiv & \!\! \int_{0}^{T}\!\!d\tau 
	\int\!\!d^{3}x  \ ,  \\
	J_{4} \!\! & \equiv  &  \!\!  iJ_{0}  \ .
\end{eqnarray}
Integrating with respect to $\Pi^{(\alpha)}$, then inserting the 
Gaussian identity, 
\begin{equation}
{\bf I}=\int\!\!{\cal D}A^{4}(\tau,{\mbit x})
[\det{(-\Nabla^{2})}]^{{1\over2}}\exp\left[
-\int\!\!d^{4}x_{_{E}}{1\over2}A^{4}(\tau,{\mbit 
x})(-\Nabla^{2})A^{4}(\tau,{\mbit x})\right] \ ,
\end{equation}
and introducing the new integration variable such that
\begin{equation}
A^{4}(\tau,{\mbit x};\phi) \equiv A^{4}(\tau,{\mbit x})
+{\sqrt{\Nabla^2(\Nabla^2\vert\widetilde{\mbit f}
(-i\Nabla)\vert^2+1)}\over\Nabla^{2}}
\stackrel{\mbox{\Large.}}{A}^{(2)}(\tau,{\mbit x})
+{1\over\Nabla^{2}}J_{4}(\tau,{\mbit x}) \ ,
\end{equation}
we obtain
\begin{eqnarray}
\!\!\!\!\!\! Z_{T}[J]\!&\!\!\!=\!\!\!&\!\int\!\!{\cal D}A^{(\alpha)}
{\cal D}A^{4}(\tau,{\mbit x};\phi)
[\det{(-\Nabla^{2})}]^{{1\over2}}
\nonumber\\
\!\!\!\!&\!\!&\!\times\exp\Biggl[-\int\!\!d^{4}x_{_{E}}\Biggl\{{1\over2}
\Bigl(\stackrel{\mbox{\Large.}}{A}^{(1)}\!\!\!(\tau,{\mbit x})\Bigr)^{2}
-{1\over2}\stackrel{\mbox{\Large.}}{A}^{(2)}\!\!\!(\tau,{\mbit x}) \Nabla^{2}
\vert\widetilde{\mbit f}(-i\Nabla)\vert^{2}\stackrel{\mbox{\Large.}}
{A}^{(2)}\!\!\!(\tau,{\mbit x})
\nonumber\\
\!\!\!\!&\!\!&\! -{1\over2}\sum_{\alpha=1}^{2}A^{(\alpha)}\!(\tau,{\mbit x})
\Nabla^{2}A^{(\alpha)}\!(\tau,{\mbit x})
-{1\over2}A^{4}(\tau,{\mbit x};\phi)\Nabla^{2}A^{4}(\tau,{\mbit x};\phi)
\\
\!\!\!\!&\!\!&\! +A^{4}(\tau,{\mbit x};\phi)
\sqrt{\Nabla^{2}(\Nabla^{2}\vert\widetilde{\mbit f}(-i\Nabla)\vert^{2}+1)}
\stackrel{\mbox{\Large.}}{A}^{(2)}\!\!\!(\tau,{\mbit x})
+J_{4}(\tau,{\mbit x})A^{4}(\tau,{\mbit x};\phi)
+{\mbit J}(\tau,{\mbit x})\!
\cdot\!({\mbit A}(\tau,{\mbit x};\phi))\Biggr\}\Biggr]  \ . 
\nonumber
\end{eqnarray}
In this way $A^{4}(\tau,{\mbit x};\phi)$ has been recovered. Now by
recalling the relations (\ref{b21}) and  (\ref{b22}), this 
becomes
\begin{eqnarray}
	Z_{T}[J] \!\! & \!=  \! & \int\!\!{\cal D}A^{(\alpha)}{\cal D}
	A^{4}(\tau,{\mbit x};\phi)
	[\det{(-\Nabla^{2})}]^{{1\over2}}  \nonumber  \\
		\!\! & \!\! & \times \exp\!\! \left[-\int\!\!d^{4}x_{_{E}}\left\{
	{1\over4}F_{\mu\nu}(\tau,{\mbit x};\phi)F_{\mu\nu}
	(\tau,{\mbit x};\phi)
	+J_{\mu}(\tau,{\mbit x})A_{\mu}(\tau,{\mbit x};\phi)
	\right\}\right] \ ,    \label{b38}
		\end{eqnarray}
where 
\begin{equation}
	F_{\mu\nu}(\tau,{\mbit x};\phi)\equiv\partial_{\mu}
	A_{\nu}(\tau,{\mbit x};\phi)
	-\partial_{\nu}A_{\mu}(\tau,{\mbit x};\phi) \ ,
	\end{equation}
and $\mu, \nu=1,2,3,4$. $A^{i}(\tau,{\mbit x};\phi)$ is now defined, 
in view of (\ref{aphi3}), by
 \begin{equation}
	A^{i}(\tau,{\mbit x};\phi) \equiv 
	n_{(1)}^{i*}(-i\Nabla)A^{(1)}(\tau, {\mbit x})
+[\delta^{ij}-\Nabla^i\widetilde f^j(-i\Nabla)]
n_{(2)}^{j*}(-i\Nabla)A^{(2)}(\tau, {\mbit x}) \ .
\end{equation}

In (\ref{b38}), almost everything is recovered but the functional measure 
which still consists of three components. To cure this, 
the gauge degree of freedom $\omega$, (\ref{b10}), 
\begin{equation}
\omega(\tau,{\mbit x})=\int\!\!d^{3}y\,{\mbit f}({\mbit x- \mbit y})\!\cdot\!
{\mbit A}(\tau,{\mbit y})=\widetilde{\mbit f}(-i\Nabla)\!\cdot\!
{\mbit A}(\tau,{\mbit x}) \ ,
\end{equation}
is revived, by means of the delta function, giving
\begin{equation}
Z_{T}[J]=\int\!\!{\cal D}A_{\mu}\delta(\widetilde{\mbit f}(-i\Nabla)
\!\cdot\!{\mbit A}(\tau,{\mbit x}))\exp\left[-\int\!\!d^{4}x_{_{E}}\left\{
{1\over4}F_{\mu\nu}F_{\mu\nu}+J_{\mu}A_{\mu}
\right\}\right] \  ,  \label{b40}
\end{equation}
where use has been made of the relation of functional measure, 
\begin{equation}
{\cal D}A^{i}={\cal D}A^{(\alpha)}{\cal D}\omega
[\det{(-\Nabla^{2})}]^{1\over2} \ ,
\end{equation}
obtained from (\ref{norm2}).

Going back the original notation
we find the covariant expression (\ref{phi1}) 
\begin{equation}
Z_{T}[J]=\int\!\!{\cal 
D}A_{\mu}\delta\!\left(\!\int\!\!d^{4}y_{_{E}}\phi_{\mu}(x-y)
A_{\mu}(y)\right)\exp\left[-\int\!\!d^{4}x_{_{E}}\left\{
{1\over4}F_{\mu\nu}F_{\mu\nu}+J_{\mu}A_{\mu}
\right\}\right]  \  . 
\label{zj}
\end{equation}
It is a straightforward task to check that the propagator is 
correctly given by (\ref{propag}). 

In the Coulomb case, (\ref{coulomb}), that is, 
$\widetilde{\mbit f}(-i\Nabla)\equiv \Nabla/\Nabla^{2}$ in (\ref{b40}),  
 a familiar expression,
\begin{equation}
Z_{T}^{\mbox{coul}}[J] =\int\!\!{\cal 
D}A_{\mu}\delta\!\left(\! \Nabla \cdot \mbit A \right) \left| \det \Bigl( - 
\Nabla^{2}\Bigr)  \right|\exp\left[-\int\!\!d^{4}x_{_{E}}\left\{
{1\over4}F_{\mu\nu}F_{\mu\nu}+J_{\mu}A_{\mu}
\right\}\right]  \  ,
\end{equation}
is obtained. Furthermore the troubles in Steinmann's approach are evaded; 
since the expression
has no singularity at all even in the Landau gauge, 
\begin{equation}
	\phi_{\mu}(x) = {\partial_{\mu} \over \dbox_{_{E}} } \ ,      \quad  
	\dbox_{_{E}} \equiv \partial_{\mu}\partial_{\mu} \ ,
\end{equation} 
and
\begin{equation}
Z_{T}^{\mbox{land}}[J]=\int\!\!{\cal 
D}A_{\mu}\delta\!\left(\! \partial_{\mu} A_{\mu} \right) \left| \det \Bigl( - 
\dbox_{_{E}} \Bigr)  \right|\exp\left[-\int\!\!d^{4}x_{_{E}}\left\{
{1\over4}F_{\mu\nu}F_{\mu\nu}+J_{\mu}A_{\mu}
\right\}\right]  \  .  
\end{equation}

\section{Proof of Gauge Independence}
In this section, gauge independence of (\ref{zj}) is proved; in other 
words, we show that any choice of $\phi^{\mu}$ leads to the same result
in the case of the free energy as well as the 
S-matrix.  To this end, let us first study how a gauge 
transformation affects the expression (\ref{zj}): the gauge transformation 
from $A_{\mu}$ to $A'_{\mu}$ is given by 
\begin{equation}
    A_{\mu}(x)\mapsto 
    A'_{\mu}(x)=A_{\mu}(x)+\partial_{\mu}\theta(x) \ .   \label{c1}
\end{equation}    
The gauge conditions are supposed as
\begin{eqnarray}
	 \int\!\!d^{4}y_{_{E}}\phi_{\mu}(x-y)A_{\mu}(y) \!\! & = & \!\! 0  \  ,
	\nonumber  \\ \noalign{\vskip -1ex}
       \hspace{30pt} & { }  & { }  \label{gcs}  \\  \noalign{\vskip -1ex}
	\int\!\!d^{4}y_{_{E}}\phi'_{\mu}(x-y)A'_{\mu}(y) \!\! & = & \!\! 0 \  ,  
	\nonumber 
	\end{eqnarray}
respectively. The second relation can be rewritten as 
\begin{equation}
 0 = \int\!\!d^{4}y_{_{E}}\phi'_{\mu}(x-y)A'_{\mu}(y) 
 = \int\!\!d^{4}y_{_{E}}\phi_{\mu}(x-y)A_{\mu}(y) 
 	+\int\!\!d^{4}y_{_{E}}\Delta\phi_{\mu}(x-y)A_{\mu}(y) +\theta(x) \ ,
\end{equation}
by use of (\ref{c1}) and (\ref{phi}), where 
\begin{equation}
	\Delta\phi_{\mu}(x) \equiv \phi'_{\mu}(x)-\phi_{\mu}(x)\ .
\end{equation}
Therefore under the gauge conditions (\ref{gcs}), $\theta(x)$ is obtained as
\begin{equation}
\theta(x)=-\int\!\!d^{4}y_{_{E}}\Delta\phi_{\mu}(x-y)A_{\mu}(y) \ . 
\label{theta}
\end{equation}

The partition function of QED, by adding the fermionic 
part to (\ref{zj}), is found as
\begin{eqnarray}
	Z[J,\overline\eta,\eta]\!&\!=\!&\!\int\!\!{\mathcal D}
	A_{\mu}{\mathcal D}\psi
	{\mathcal D}\overline\psi\delta\left(\int\!\!d^{4}
	y_{_{E}}\phi_{\mu}(x-y)
	A_{\mu}(y) \right)
	\nonumber  \\
	\!&\!\!&\!\times\exp\left[-\int\!\!d^{4}x_{_{E}}\left\{
	{1\over4}F_{\mu\nu}F_{\mu\nu}+\overline\psi(\dsl{D}+m)\psi
	+J_{\mu}A_{\mu}+\overline\eta\psi+\overline\psi\eta
	\right\}\right] \ ,  \label{c6}
\end{eqnarray}
where the periodic boundary condition for $\phi_{\mu}$ and the
 anti-periodic boundary condition for fermions must be 
understood\footnote{The continuum representation for fermion is 
problematic\cite{KSW}. But here we concentrate ourselves on perturbation 
theories so that we neglect the Wilson term etc. .}. Meanwhile the transformed partition function is 
\begin{eqnarray} 
	Z'[J,\overline\eta,\eta]\!&\!=\!&\!\int\!\!{\mathcal D}A'_{\mu}
	{\mathcal D}\psi'{\mathcal D}\overline\psi'
	\delta\left( \int\!\!d^{4}y_{_{E}}\phi'_{\mu}(x-y)A'_{\mu}(y) \right)
	\nonumber  \\
	\!&\!\!&\!\times\exp\left[-\int\!\!d^{4}x_{_{E}}\left\{
	{1\over4}F'_{\mu\nu}F'_{\mu\nu}+\overline\psi'(\dsl{D}'+m)\psi'
	+J_{\mu}A'_{\mu}
	+\overline\eta\psi'
	+\overline\psi'\eta
	\right\}\right]  \  , 
\end{eqnarray}
where $A'_{\mu}$ has been given in (\ref{c1}) and
\begin{equation}
  \psi(x) \mapsto \psi'(x) \equiv e^{i\theta(x)} \psi(x)  \ , 
  \hspace{2ex} \overline\psi(x) 
  \mapsto  \overline{\psi'}(x)  \equiv \overline\psi(x)  e^{-i\theta(x)} 	 \ .
\end{equation}
Then a simple change of variables with a trivial Jacobian\footnote{Since 
the Jacobian from 
$A'_{\mu}$ to $A_{\mu}$ reads 
$\left|{\delta A'_{\mu}\over\delta A_{\nu}}\right|
=|\det(\delta_{\mu\nu}-\partial_{\mu}\Delta\phi_{\nu})|$, 
according to (\ref{c1}) 
and (\ref{theta}), which is unity: consider the determinant 
of the matrix, 
$M_{ij}=\delta_{ij}+A_{i}B_{j}$ with $\sum_{i}A_{i}B_{i}=0$ to find $\det M=
\exp[{\mathrm Tr}\log(1+AB)] =1$; since ${\mathrm Tr} (AB)^{n}=0; \ 
{}^{\forall } n$.} leads to
\begin{eqnarray}
Z'[J,\overline\eta,\eta]\!&\!=\!&\!\int\!\!{\mathcal D}A_{\mu}{\mathcal D}\psi
	{\mathcal D}\overline\psi\delta 
	\left(\int\!\!d^{4}y_{_{E}} \phi_{\mu}(x-y)
	A_{\mu}(y) \right)
	\nonumber  \\
	\!&\!\!&\!\times\exp\left[-\int\!\!d^{4}x_{_{E}}\left\{
	{1\over4}F_{\mu\nu}F_{\mu\nu}+\overline\psi(\dsl{D}+m)\psi
	+J_{\mu}\left(A_{\mu}+\partial_{\mu}\theta\right)
	+\overline\eta\mathrm{e}^{ie\theta}\psi
	+\overline\psi\mathrm{e}^{-ie\theta}\eta
	\right\}\right] \  .
\end{eqnarray}
Therefore if the sources, $J,\overline\eta,$ and $\eta$, are absent, the 
relation
\begin{equation}
Z' = Z \left(=  {\mathrm Tr}e^{-T H} \right) \  ,
\end{equation}
implies that the free energy of QED is gauge independent.  Moreover, 
it can be recognized that expectation values of a gauge invariant 
operator, such as the Belinfante's energy-momentum tensor, 
$\bra{ \{ n \} } {\mit \Theta}_{\mu\nu}(x) \ket{ \{n\} }$ is gauge 
invariant\footnote{(\ref{c6}) itself is gauge invariant without 
sources for an arbitrary $T$, that is, ${\rm 
Tr}{\mit \Theta}_{\mu\nu}(x) e^{-TH}= 
\sum_{\{n\}}\bra{ \{n\} } {\mit \Theta}_{\mu\nu}(x) \ket{ \{n\} }$ is gauge 
invariant. Then, $T \rightarrow \infty$ picks up the expectation 
value between the vacuum $\ket{0}$; which is gauge invariant. Next, put 
$T \rightarrow \infty$ in the quantity, 
$Z[J=0,\overline\eta=0,\eta=0]-\bra{0}{\mit \Theta}_{\mu\nu}(x) \ket{ 0 }$, 
to give the expectation value between the first excited state; which is 
again gauge invariant. Repeating the procedures, we find the expectation 
value between any state is gauge invariant. q.e.d. }, 
where $\ket{\{n\}}$ designates states of electrons and photons, and 
\begin{eqnarray}
	 {\mit \Theta}_{\mu\nu} \!\! & \equiv & \!\! {i \over 4} 
	 \overline{\psi} \left(
\gamma_\mu
\lrDer{\nu}{}  + \gamma_\nu \lrDer{\mu}{} \right) \psi- F_{\mu
\rho}{F_\nu}^\rho -g_{\mu \nu} {\cal L}  \ ,    \nonumber
	 \\   \noalign{\vskip -1ex}
	\!\! & \!\!& \!\!            
	\\  \noalign{\vskip -1ex}
{\cal L}  \!\! & \equiv & \!\! \overline{\psi} \left({i\over2}\lrDerb  -m \right)
\psi - {1 \over 4} F_{\mu \nu} F^{\mu \nu}  ,  \  \nonumber 
\end{eqnarray}
with 
\begin{equation}
\overline{\psi} \lrDer{\mu}{} \psi
\equiv \overline{\psi} \Bigl( ( \der{\mu}{} -ieA_\mu )\psi \Bigr) 
 - \Bigl( (  \der{\mu}{} + ieA_\mu  ) \overline{\psi}  
 \Bigr)\psi   .  \
\end{equation}
(We have employed the Minkowski metric, here.)
In this way, the path integral gives us a more quick and intuitive derivation of 
gauge independence, whose reason is clarified in the next section.

In order to discuss a gauge independence of the S-matrix, however, we 
need a further consideration: suppose $\theta(x)$ is infinitesimal so that the
difference between 
$Z'[J,\overline\eta,\eta]$ and $Z[J,\overline\eta,\eta]$ is 
\begin{eqnarray} 
	\Delta Z[J,\overline\eta,\eta]
	\!&\!=\!&\!Z'[J,\overline\eta,\eta]-Z[J,\overline\eta,\eta]
	\nonumber \\
	\!&\!=\!&\!\int\!\!{\mathcal D}A_{\mu}
	{\mathcal D}\psi{\mathcal D}\overline\psi
	\delta\left(\int\!\!d^{4}y_{_{E}}\phi_{\mu}(x-y)A_{\mu}(y)\right)
	\int\!\!d^{4}x_{_{E}}\left[\theta\partial_{\mu}J_{\mu}
	+ie\theta\left(\overline\psi\eta-\overline\eta\psi\right)\right]
	\nonumber  \\
	\!&\!\!&\!\times\exp\left[-\int\!\!d^{4}x_{_{E}}\left\{
	{1\over4}F_{\mu\nu}F_{\mu\nu}+\overline\psi(\dsl{D}+m)\psi
	+J_{\mu}A_{\mu}
	+\overline\eta\psi
	+\overline\psi\eta
	\right\}\right] \  
 	\label{diff}
\end{eqnarray}
which is the generating functional of the Green's function
 \begin{equation}
	 \Delta 
	G^{(n)}\equiv G'^{(n)}-G^{(n)} \ , 
 \end{equation}
with $G'^{(n)}$ and $G^{(n)}$ 
being the n-point Green's functions in each gauge. 

The S-matrix is now found, after rotating back to the Minkowski space, 
by cutting external legs and multiplying the wave functions of
electrons and photons, that is, multiplying
\begin{eqnarray}
 \!\! & \!\! & \!\! {\dslp-m\over i\sqrt{z_{2}} }
	{u({\mbit p},s)\over\sqrt{(2\pi)^{3}2p_{0}}} \ ,  
	{ { \overline u({\mbit p},s)}  \over\sqrt{(2\pi)^{3}2p_{0}} }
	{\dslp-m\over i\sqrt{z_{2}}} \ ,  \quad \mbox{for electrons} 
	\nonumber \\
	\noalign{\vskip -1ex}
	\!\! & \!\! & \!\!       \label{asympt} \\
	\noalign{\vskip -1ex}
       \!\! & \!\! & \!\!   \hspace{10ex}	{-q^{2}\over i\sqrt{z_{3}} }
	{\xi_{\mu}^{(i)}({\mbit q})\over\sqrt{(2\pi)^{3}2q_{0}}} \ ,  
	\hspace{14ex}
	\mbox{for photons}  \nonumber 
\end{eqnarray}
to the Green's function\footnote{It is troublesome to write out the 
LSZ-asymptotic state for electrons in this way; since in a noncovariant 
gauge $z_{2}$ is given as matrix-valued acting differently on each 
spinor index. However, there are additional renormalization conditions, 
since the self-energy is not merely the function of 
$\dslp$: it depends on $p_{0}\gamma_{0}$ as well as 
$p_{k}\gamma_{k}$ in the Coulomb gauge for instance. Here we assume
that $z_{2}$ has already been diagonalized by utilizing these additional degrees.}. 
Here the photon polarization $\xi_{\mu}^{(i)}(\mbit q)$ 
fulfills  the transversal condition $q^{\mu}\xi_{\mu}^{(i)}(\mbit q)=0$. 
Due to this, $\theta\partial_{\mu}J_{\mu}$ term in (\ref{diff}) (now 
$\theta\partial^{\mu}J_{\mu}$) drops 
out so that it is enough to concentrate on differences of the electron 
legs: the gauge dependent part of the S-matrix, $S_{g}$, is thus read
\begin{equation}
S_{g} = \prod_{j=1}^{n}
{\overline u({\mbit p}_{j},s'_{j})\over\sqrt{(2\pi)^{3}2(p_{j})_{0}}}
	{\dslp_{j}-m\over i\sqrt{z_{2}}}
G^{(2n)}(p_{1},\cdots,p_{n};k_{1},\cdots,k_{n})
\prod_{j=1}^{n}{\dslk_{j}-m\over i\sqrt{z_{2}}}
	{u({\mbit k}_{j},s_{j})\over\sqrt{(2\pi)^{3}2(k_{j})_{0}}} \ .  \label{es}
\end{equation}

\begin{figure}[t]
\centering
\includegraphics*[scale=.5]{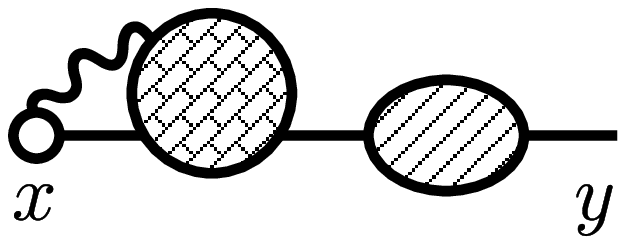}
\hspace{1ex}\raisebox{3ex}{$-$}\hspace{1ex} 
\includegraphics*[scale=.5]{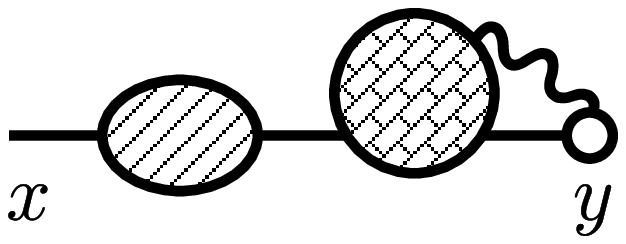}
\raisebox{3ex}{;} \hspace{2ex} \raisebox{3ex}{where}\hspace{1ex}
\includegraphics*[scale=.4]{fig1a.eps} 
\ \raisebox{3ex}{$\equiv$}\ 
\includegraphics*[scale=.4]{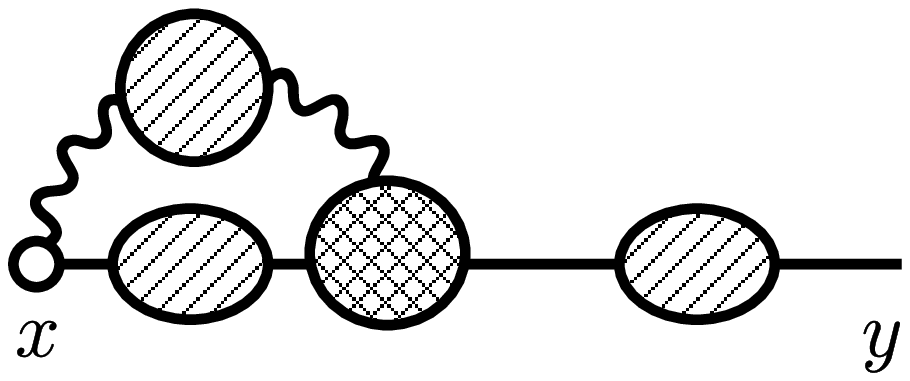} 
\caption{The two-point function in (\ref{c18}): the circle denotes the 
$\theta$-insertion. The blob in the left graphs is 
collections of the full propagators and the vertex, seen in the right 
graph.}
\label{FirstFigure}
\end{figure}

Consider the Fourier transform of the difference of $2n$-point function:
\begin{eqnarray}
	\lefteqn{\Delta G^{(2n)}(p_{1},\cdots,p_{n};k_{1},\cdots,k_{n})
	\bigg\vert_{k_{n}=p_{n}+\sum_{j=1}^{n}(p_{j}-k_{j})}}  \nonumber \\
	\!&\! \equiv \!&\!
	\int\!\!\prod_{j=1}^{n-1}(d^{4}x_{j}d^{4}y_{j})d^{4}x_{n}
	\exp{\left[\sum_{j=1}^{n-1}(ip_{j}x_{j}-ik_{j}y_{j})+ip_{n}x_{n}
	\right]}
\nonumber \\
	\!&\!\!&\!   \left. \hspace{15ex} \  \times \   
	{\delta^{2n}\over{\delta\overline\eta(x_{1})\cdots\delta\overline\eta(x_{n})
           \delta\eta(y_{1})\cdots\delta\eta(y_{n-1})\delta\eta(0)}}
           \Delta Z[J,\overline\eta,\eta] \right|_{J=\overline\eta=\eta=0}
\label{c16} \\
	\!&\!=\!&\!\int\!\!\prod_{j=1}^{n-1}(d^{4}x_{j}d^{4}y_{j})d^{4}x_{n}
	\exp{\left[\sum_{j=1}^{n-1}(ip_{j}x_{j}-ik_{j}y_{j})+ip_{n}x_{n}
	\right]}
\nonumber \\
	\!&\!\!&\! \times  \  ie \bra{0}{\mathrm T}
	\biggl\{\sum_{j=1}^{n-1}[\theta(x_{j})-\theta(y_{j})]
	+[\theta(x_{n})-\theta(0)]\biggr\}
	\psi(x_{1})\cdots\psi(x_{n})\overline\psi(y_{1})\cdots\overline
	\psi(y_{n-1})\overline\psi(0)\ket{0}\ .   \nonumber
\end{eqnarray}
In view of (\ref{theta}), $\theta$ contains $A_{\mu}$ and cannot be put 
outside of the expectation value. The two-point function, that is, $n=1$ case,
\begin{equation}
	\Delta G^{(2)}(x,y) =  ie\bra{0}{\mathrm T}(\theta(x)-\theta(y))
    \psi(x)\overline\psi(y)\ket{0} \ ,  \label{c18}
\end{equation}
depicted in figure \ref{FirstFigure},

becomes near the mass-shell such that
\begin{equation}
	\Delta G^{(2)}(p)\sim\overline A(p)\biggr\vert_{\dslp=m}
	{iz_{2}\over\dslp-m}
	-{iz_{2}\over\dslp-m}A(p)\biggr\vert_{\dslp=m}
               \ , \label{diffg2}
\end{equation}
where 
\begin{eqnarray}
	\overline A(p)  \!\! & \stackrel{F.T.}{=}& \!\! \int\!\! d^{4}z \ 
	 ie \bra{0}{\mathrm T}\theta(x)
    \psi(x)\overline\psi(z)\ket{0}(G^{(2)})^{-1}(z,y) \  ,
	\nonumber \\
	\noalign{\vskip -1ex}
	\!\! & \!\! & \!\!
	\\
	\noalign{\vskip -1ex}
	A(p) \!\! & \stackrel{F.T.}{=}& \!\! \int\!\! d^{4}z \ 
	 ie (G^{(2)})^{-1}(x, z) \bra{0}{\mathrm T}\theta(y)
    \psi(z)\overline\psi(y)\ket{0}    \ ,
	\nonumber
\end{eqnarray}
and $F.T.$ designates the Fourier transformation; since the electron 
two-point function behaves  
\begin{equation}
	G^{(2)}(p)\sim{iz_{2}\over\dslp-m} \   ,
\end{equation}
near the mass-shell. There is apparently no poles  in $\overline A(p)$ and $A(p)$
when $\dslp=m$ so that the right hand side of (\ref{diffg2}) is 
written as 
\begin{equation}
  	\Delta G^{(2)}(p)\sim	{i \Delta z_{2} \over\dslp-m} \  
	 \label{c22}
\end{equation}
implying that the gauge difference reads as the change of wave function 
renormalization constant $z_{2}$: 
\begin{equation}
	z'_{2}=z_{2}+\Delta z_{2} \  . 
 	\label{dz}
\end{equation}

\begin{figure}[tbp]
	\centering
	\includegraphics*[scale=.5]{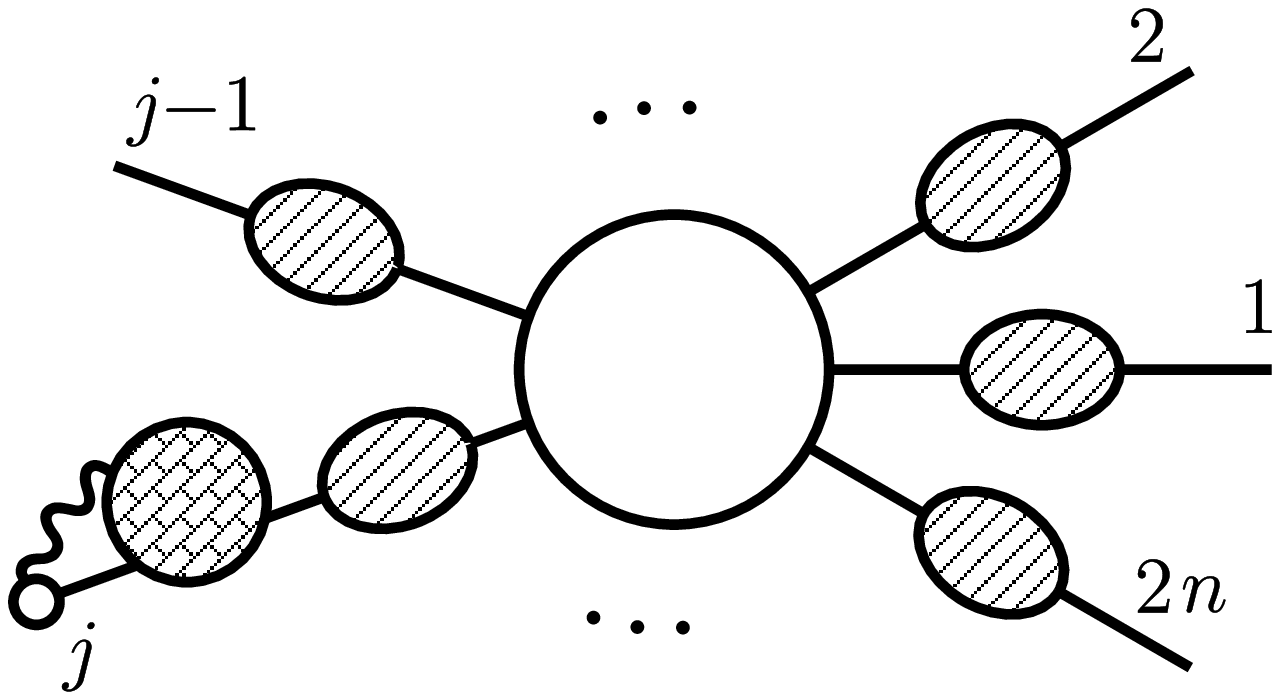} \hspace{8ex} 
	\includegraphics*[scale=.5]{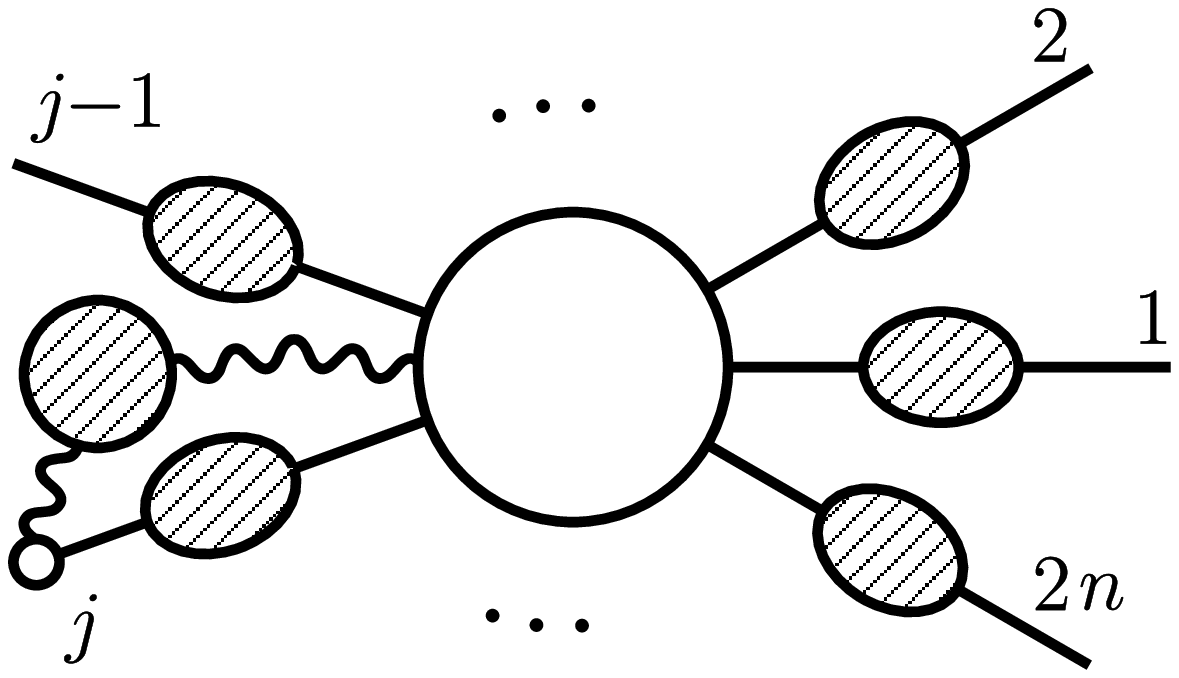}
	\caption{Left: Graphs that do contribute. Right: those do not 
	contribute. Big circles at the center of each graph denote 
	the amputated Green's functions.}
	\label{SecondFigure}
\end{figure}

Due to $(\dslp_{i}-m)$ or $(\dslk_{i}-m)$ in (\ref{es}), the 
surviving part of $\Delta G^{(2n)}$ must have
$2n$ one-particle poles. Graphs (see the figures \ref{SecondFigure}), in which the photon in $\theta$ 
(\ref{theta}) is attached to its original electron line, that is, 
graphs including  
$\overline A(p)$ and $A(p)$, have the same 
pole structure as $G^{(2n)}$ and do contribute, 
but those, in which the photon goes somewhere other than its original electron 
line change the pole structure then do not contribute to (\ref{dg}). 
Therefore write the former as $\Delta\overline G^{(2n)}$ 
to find
\begin{eqnarray}
\lefteqn{\Delta\overline G^{(2n)}(p_{1},\cdots,p_{n};k_{1},\cdots,k_{n})}
\nonumber     \\
\!&\!=\!&\!\sum_{j=1}^{n}\left(\overline A(p_{j})\bigg\vert_{\dslp_{j}=m}
G^{(2n)}(p_{1},\cdots,p_{n};k_{1},\cdots,k_{n})
-G^{(2n)}(p_{1},\cdots,p_{n};k_{1},\cdots,k_{n})
A(k_{j})\bigg\vert_{\dslk_{j}=m}\right)
\label{dg}     \\
\!&\!=\!&\!\sum_{j=1}^{n}{\Delta z_{2}\over z_{2}}
G^{(2n)}(p_{1},\cdots,p_{n};k_{1},\cdots,k_{n})
={n\Delta z_{2}\over z_{2}}G^{(2n)}(p_{1},\cdots,p_{n};k_{1},\cdots,k_{n})\ .
\nonumber 
\end{eqnarray}

The difference of the gauge dependent part, $S_{g}$ (\ref{es}), is given by
\begin{eqnarray}
	\Delta S_{g} \!&\! \equiv \!&\!S'_{g}-S_{g}
	\nonumber \\
	\!&\!=\!&\!\prod_{j=1}^{n}
	{\overline u({\mbit p}_{j},s'_{j})\over\sqrt{(2\pi)^{3}2(p_{j})_{0}}}
	{\dslp_{j}-m\over i\sqrt{z_{2}}}
	\Delta\overline G^{(2n)}(p_{1},\cdots,p_{n};k_{1},\cdots,k_{n})
	\prod_{j=1}^{n}{\dslk_{j}-m\over i\sqrt{z_{2}}}
	{u({\mbit k}_{j},s_{j})\over\sqrt{(2\pi)^{3}2(k_{j})_{0}}}
	 \\
	\!&\!\!&\!\hspace{20pt}+\Delta z_{2}{\partial\over\partial z_{2}}
	z_{2}^{-n}\cdot(z_{2}^{n} S_{g})
\nonumber
\end{eqnarray}
where the first term comes from the change of $G^{(2n)}$ and the 
second term comes from the change of $z_{2}$ (\ref{dz}). Due to 
(\ref{dg}), it is apparent that $\Delta S_{g}$ vanishes. There is no 
gauge dependence in the S-matrix.

\section{Functional Method as an Efficient Tool for Handling Gauge 
Theories}

In this section we make a detailed discussion why we can perform a 
gauge transformation so easily and intuitively in the functional representation. 
As was mentioned in the introduction, gauge transformation in the
canonical operator formalism cannot be allowed at all. In this sense, 
it is instructive to study the 
$A_0 =0$ gauge in the conventional treatment\cite{JSCL}; since there 
needs a supplementary condition, so called a physical state condition,
 implying that {\em physical state must be
gauge invariant.} The statement is apparently contradict with the 
above situation. 

In $A_0 =0$ gauge, all three components ${\mbit A}$ are assumed dynamical and 
obey the commutation relations,
\begin{equation}
[ \hat{A}_j({\mbit x}), \hat{\Pi}_k({\mbit y}) ] = i\delta_{jk} \delta({\mbit x} - 
 {\mbit y}) ,
\quad    [\hat{A}_j({\mbit x}), \hat{A}_k({\mbit y})] = 0 = [\hat{\Pi}_j({\mbit x}),
\hat{\Pi}_k({\mbit y})] ; \quad (j, k = 1, 2, 3) \  .
\end{equation}
Again the caret designates operators. The physical state condition 
is given as
\begin{equation}
\hat{\Phi}({\mbit x})| \mbox{phys}\rangle \equiv
\bigg[ \sum_{k=1}^3 \big(
\partial_k \hat{{\mbit \Pi}}_k({\mbit x}) \big) + J_0({\mbit x}) 
\bigg] | \mbox{phys} \rangle =
0 \ ,    \label{d2}
\end{equation}
where $J_\mu(x)$ is supposed as a $c$-number current.  First this should 
be read such that {\em there is no gauge transformation in the physical space}: 
$\hat{\Phi}$ is the generator of the gauge transformation. However, 
the representation cannot be obtained within the usual Fock 
space; since
$\hat{\Phi}({\mbit x})$ is a local operator to result in 
$\hat{\Phi}({\mbit x})=0$\cite{PR}, which is another consequence of the 
above statement.  Nevertheless, the state can be expressed in the functional 
(Schr\"odinger) representation\cite{FJ}:
\begin{eqnarray}
\widehat{\mbit A}({\mbit x}) | \{{\mbit A} \} \rangle \!\! & =  & 
\!\!  {\mbit A} ({\mbit x}) | \{{\mbit A} \} \rangle  \  ,  \quad 
\widehat{\mbit \Pi } ({\mbit x})  | \{{\mbit \Pi} \} \rangle = 
 {\mbit \Pi} ({\mbit x}) | \{{\mbit \Pi} \}
\rangle \ ,  \nonumber 	 \\     
 \noalign{\vskip -1ex} 
\!\! & \!\! & \!\!      \\      
   \noalign{\vskip -1ex}        
   \langle \{ {\mbit A} \} | \widehat{ \mbit \Pi} ({\mbit x}) 
 \!\! &  =  & \!\!    
  -i { \delta  \over {\delta  {\mbit A}({\mbit x}) } }\langle \{ {\mbit A} \} | 
   \  ,                \quad   \dots      \nonumber    
\end{eqnarray}

To see the reason consider the state, $| \{{\mbit A} \} \rangle$, which can
be constructed in terms of the Fock states as follows: the creation and annihilation 
operators are given by 
\begin{eqnarray}
	\!\!  & \!\! & \!\! \widehat{\mbit A}({\mbit x})  =  
	 \int {d^3 {\mbit k} \over {(2 \pi)^{3/2} \sqrt {2|
{\mbit k}| } }} \big( {\bf a}( {\mbit k} ) e^{i{\mbit k} \cdot {\mbit x} }  
 +   {\bf a}^{\dag} (
{\mbit k} ) e^{- i{\mbit k} \cdot {\mbit x} } \big) 
	     \nonumber  \\  
	     \noalign{\vskip -1ex}  
\!\!  & \!\! & \!\!          \label{expan}  \\
\noalign{\vskip -1ex} 
	\!\!  & \!\! & \!\!  [a_i({\mbit k} ),   a_j^{\dag}({\mbit k}')] =
\delta_{ij}\delta({\mbit k}-{\mbit k}'),
\quad  [a_i({\mbit k} ), a_j({\mbit k}')] = 0 \ ,
	\nonumber
\end{eqnarray}
and the vacuum $|0 \rangle$ obeys
\begin{equation}
 	{\bf a}( {\mbit k} ) |0 \rangle = 0 \ . \label{vacuum}
 \end{equation} 
 Now recall the quantum mechanical case\cite{KAS}:
\begin{eqnarray}
\!\! & \!  \! & \!\!   \hspace{8ex}	\hat q | q \rangle =  q  | q \rangle \ ,  \qquad  \quad
\hat p | p \rangle = p | p
\rangle \ ,	\nonumber   \\
\noalign{\vskip -1ex}
	\!\! & \!\! & \!\!
	\label{qprep}  \\ 
	\noalign{\vskip -1ex}
\!\! & \! \! & \!\!  \hat q	= { 1 \over \sqrt2}  \left( a  + a^{\dag} \right) \ , 
\quad \hat p = { 1
\over {\sqrt2}i} \left( a -  a^{\dag} \right) \  ;  \qquad  a |0\rangle =0 \  , 
\nonumber
\end{eqnarray}
then
\begin{eqnarray}
   \!\!	 & \!\! &  \!\! | q \rangle = {1 \over \pi^{1/4} }
 \exp \left( - { q^2 \over 2} + \sqrt2 q a^{\dag} - {( a^{\dag})^2 \over 2} \right)  | 0
\rangle \  ,  \nonumber \\
      \noalign{\vskip -1ex}
	\!\!  & \!\!  & \!\! 
	\label{qprepa}  \\
	\noalign{\vskip -1ex}
  \!\! & \!\! & \!\!  | p \rangle  = {1 \over \pi^{1/4} } \exp \left( - { p^2 \over 2} 
  + \sqrt 2 ip
a^{\dag} + {  ( a^{\dag} )^2 \over 2 } \right)  | 0 \rangle \  . 
       \nonumber
\end{eqnarray}
These bring us to the expression:
\begin{eqnarray}
	      | \{{\mbit A} \} \rangle  &
\underline{\sim }  &  \exp \Big[ -{1 \over 2} \int d^3 {\mbit x} \ d^3{\mbit y}
 \  {\mbit A}( {\mbit x} ) K( {\mbit x} - {\mbit y} ) {\mbit A}({\mbit y} )
	\nonumber   \\
		\noalign{\vskip -1ex}
	\!\!& \!\!  & \!\!
	\label{arep}  \\
	\noalign{\vskip -1ex}
	\!\!  & +  & \!\! \int d^3 {\mbit x}
\! \!  \int d^3 { \mbit  k}  \  
\sqrt{ 2|{\mbit k} |  \over  (2\pi )^3  }  \ {\mbit A} ( {\mbit x} ) \! \cdot \! 
{\bf a}^{\dag} ({\mbit k}
) e^{-i{\mbit k} \cdot {\mbit x} }   
 - {1 \over 2} \int d^3 {\mbit k}  \  {\bf a}^{\dag} ({\mbit k} ) \! \cdot \!  
  {\bf a}^{\dag} (-
{\mbit k} ) \Big] | 0 \rangle  \  , 
	\nonumber
\end{eqnarray}
where 
\begin{equation}
K({\mbit x}) \equiv \int  { d^3 {\mbit k} \over  (2\pi )^3 }  | {\mbit k} | 
e^{i {\mbit k} \cdot {\mbit x} }
 \ ,  	
	\label{kfun}
\end{equation}
which is apparently divergent:
\begin{equation}
K({\mbit x})    = 	O\Bigl({ \Lambda^{2} \over | {\mbit x} | 
}\Bigr)  \ , 
\end{equation}
where $\Lambda$ is some cut-off.  
The physical state in the functional
representation is thus found as
\begin{equation}
\langle \{{\mbit A} \} | \hat{\Phi}({\mbit x})  |  \mbox{phys} \rangle  = \left(
-i {\Nabla}  {\delta \over  {\delta  {\mbit A}({\mbit x}) } } 
 - J_0({\mbit x}) \right)
\Psi_{\rm phys}[{\mbit A}] = 0 \ , 	
	\label{phystasol}
\end{equation}
where
\begin{equation}
\Psi_{\rm phys}[{\mbit A}]  \equiv \langle \{{\mbit A} \} 
|   \mbox{phys} \rangle \ .
\end{equation}

Therefore physical state can be obtained under the functional 
representation, implying that gauge transformations are permissible. 
Now we should see the reason: within a single Fock state the
physical state condition (\ref{d2}) merely implies $\hat{\Phi}({\mbit 
x})=0$. However, we should bear the following fact in mind: 
{\em the functional representation consists of infinitely many 
collections of inequivalent Fock spaces}; since the inner
product of $| \{{\mbit A} \} \rangle$ 
(\ref{arep}) to the Fock vacuum is found to be 
\begin{eqnarray}
	\langle \{{\mbit A} \}  | 0 \rangle  \!\! &  \sim&  \!\! \exp\left[  
- { 1 \over 2} \int d^3
{\mbit x} \ d^3{\mbit y} \ {\mbit A}( {\mbit x} ) \ K( {\mbit x} - {\mbit y} ) 
{\mbit  A}({\mbit y} )
\right]
	\nonumber  \\
	\noalign{\vskip -1ex}
	\!\! & \!\!  & \!\! 
		\label{inner}  \\
	\noalign{\vskip -1ex}
\!\!  & \sim  & \!\! \exp\left[  - { \Lambda^{2} \over 2} \int d^3  {\mbit x}   
d^3{\mbit  y } 
{ {\mbit A}( {\mbit x} ) {\mbit  A}({\mbit 
y} ) \over | {\mbit  x} - {\mbit  y} | } \right]
\stackrel{\Lambda \rightarrow \infty}{\longrightarrow}   0
 \    \ .  
	\nonumber 
\end{eqnarray}
Note that in the functional space ${\mbit A}( {\mbit x} )$ is far from
a Fourier transformable form\footnote{Recall that even the free 
theory the action is divergent, implying those do not belong to 
${\cal L}^{2}$ class.}. Therefore the space of ${\mbit A}( {\mbit x} )$ 
which makes the exponent in the last relation finite is almost 
measure zero. Then we can say that (\ref{inner}) happens in 
{\em any value of} ${\mbit A}({\mbit x})$. Thus the functional 
representation for {\em any} ${\mbit A}({\mbit x})$ is orthogonal to 
the Fock state, that is, inequivalent to the Fock state. Any local first class constraint, 
(apart from the mathematical rigorousness of that),
can be realized by means of the functional
representation.

This fact that the functional representation 
contains an infinite set of the Fock states enables us to perform an explicit gauge
transformation and prove gauge independence without recourse to any physical
state conditions in path integral. (Recall that (\ref{b26}) and 
(\ref{b27}) are the essential ingredients toward the path integral 
representation.)

\section{Discussion}

In this paper, we have built up the path integral formula of the abelian gauge 
theory in which an arbitrary gauge function is included. Although in the 
operator formalism the support of the function must be in a spacelike 
region, thus generality is lost, there is no restriction in the Euclidean
path integral  expression so as to be able to move, for instance, 
from the Coulomb to the 
Landau gauge.  By utilizing arbitrariness of the function, gauge independence 
of the free energy is quickly understood (but that of the S-matrix 
needs further considerations.)
Furthermore, a closer inspection reveals the reason why 
gauge transformations are so easily managed in the path integral.

As was seen in the discussion of the S-matrix, multiplying wave 
functions, that is, the on-shell condition, is indispensable for the 
proof of gauge independence. The on-shell condition belongs to one of the 
physical state conditions. Hence, in scattering theories or in perturbation
theories, usual (LSZ-)asymptotic states\cite{NN}, (\ref{asympt}), are 
known to behave as the physical states. 
However, it is not so easy to find out the form of the physical state 
in a nonperturbative manner.  Steinmann's first intention seems to explore
this: indeed, the 
physical electron (\ref{psi}) by taking $\phi_{\mu}$ as the Coulomb case 
(\ref{coulomb}), 
\begin{equation}
	\Psi^{\rm D}(x) = \exp\left[-ie {\Nabla \cdot {\mit A} \over \Nabla^{2}}
	\right]\psi(x)   \ , 
\end{equation}
is the one introduced by Dirac\cite{DT}, which is locally gauge 
invariant as well as globally charged. The existence of such a state 
implies an evidence of electron as a real particle\cite{LM}. 

The issue should then be raised to 
the nonabelian gauge case (QCD). In order to study dynamics of the quark
confinement, it is important to examine whether
physical charged state can be constructed or not. The key to this 
direction would be to notice the Gribov ambiguity\cite{GA}: in a 
smaller region, the Coulomb gauge is well-defined, that is, no gauge 
degrees of freedom being left owing to the 
asymptotic freedom. The larger a region, however, the more nontrivial 
degree comes into a part\cite{LMc}. Since gauge invariance is 
essential to comprehend the quark confinement, the path integral must 
be useful. Therefore in order for the theory 
to be well-defined in terms of the path integral the integration region of 
the gauge fields must pertain to that of the Lagrangian, which would 
finally give us a compact integration of gauge fields given by the 
lattice QCD\cite{WC}. A 
work in this direction is in progress.

\vspace{5ex}  
\centerline{\bf Acknowledgment}
\vspace{2ex}
\noindent The authors thank to I. Ojima for guiding them to the work of 
Steinmann and to K. Harada for discussions.


\begin{thebibliography}{99}
\bibitem{PTF}
E.A.Power and T.Thirunamachandran, Am.J.Phys.{\bf 46}(1978)380.\\
J.J.Forney, A.Quattropani, and F.Bassani, Nuovo Cimento {\bf 37B}(1977)78.\\

\bibitem{BB}
I. Bialynicki-Birula, Phys. Rev.  {\bf D2} (1970) 2877.  \\
B.W. Lee and J. Zinn-Justin, Phys. Rev. {\bf D5} (1972) 3121, 3137. \\
G. 't Hoot and M. Veltman, Nucl. Phys. {\bf B50} (1972) 318.


\bibitem{KFC}
E.Kazes, T.E.Feuchtwang, P.H.Cutler, and H.Grotch, Ann.Phys.{\bf 
142}(1982)80.\\
K.Haller and E.Lim-Lombridas, Found. Phys.{\bf 24}(1994)217.\\

\bibitem{NL} 
N. Nakanishi,
Prog. Theor. Phys. {\bf 35} (1966) 1111; {\bf 49} (1973) 640.\\
B. Lautrup, Mat. Fys. Medd. Dan. Vid. Selsk. {\bf 35} (1967) 29. 
%
\bibitem{BNS}
A. Bassetto, G. Nardelli, and R. Soldati, ``YANG-MILLS THEORIES IN 
ALGEBRAIC NON-COVARIANT GAUGES'' World Sientific, 1991. 


\bibitem{ST}
O. Steinmann, Ann. Phys. {\bf 157} (1984) 232.
%
\bibitem{KaTa}
T. Kashiwa and N. Tanimura, {\it Physical States and Gauge 
Independence of the Energy-Momentum Tensor in Quantum Electrodynamics}
(hepth-9605207, KYUSHU-HET-31, May 1996) 

\bibitem{AL}
E.~S.~Abers and B.~W.~Lee, Phys. Rep. {\bf 9c} (1973) 1. 

\bibitem{KS}
T. Kashiwa and M. Sakamoto, Prog. Theor. Phys. {\bf 67} (1982) 1927. \\
Also see T. Kashiwa, Prog. Theor. Phys. {\bf 66 } (1981)  1858. 
%

%
\bibitem{FS}
L. D. Faddeev and A. A. Slavnov, ``Gauge Fields,'' chap.3,  Benjamin, Inc. 1980.
%
\bibitem{Tab}
In the Coulomb case, see Y. Takahashi, Physica, {\bf 31} (1965) 205.


\bibitem{KSW}
T. Kashiwa and H. So, Prog. Theor. Phys. {\bf 73} (1985) 762. \\
K. G. Wilson, in {\em New Phenomena in Subnuclear Physics}, Proc. 14th 
Int. School of Subnuclear Physics, Erice 1975, ed. A. Zichichi 
(Plenum Press, New York, 1977).


\bibitem{JSCL}
J. L. Gervais and B. Sakita, Phys. Rev. {\bf D18} (1978) 453. \\
N. H. Christ and T. D. Lee, Phys. Rev. {\bf D22} (1980) 939.
%
\bibitem{PR}
P. G. Federbush and K. A. Johnson, Phys. Rev. {\bf 120} (1960) 1926. \\
P. Roman, ``Introduction to Quantum Field Theory,'' p.381, John Wiley \& Sons,
Inc. 1969. 


\bibitem{FJ}
R. Floreanini and R. Jackiw, Phys. Rev. {\bf 37} (1988) 2206. 
%

\bibitem{KAS}
T. Kashiwa, Prog. Theor. Phys. {\bf 70} (1983) 1124. 
%
\bibitem{DT}
P. A. M. Dirac, ``Principle of Quantum Mechanics,''  p. 302, Oxford University Press, Oxford,
1958.  \\
Also see
T.~Kashiwa and Y.~Takahashi, {\it Gauge Invariance in Quantum Electrodynamics}
(KYUSHU-HET-14, January 1994) unpublished. 
%

\bibitem{NN}
N. Nakanishi, Prog. Theor. Phys. {\bf 52} (1974) 1929.

\bibitem{LM} 
M. Lavelle and D. McMullan, Phys. Rev. Lett. {\bf 71} (1993) 3758. \ 
Phys. Lett. {\bf 312B}  (1993) 211. 
%
%
\bibitem{GA}
H. D. I. Abarbanel and J. Bartels, Nucl. Phys. {\bf 136} (1978) 237 .\\
V. N. Gribov, Nucl. Phys. {bf 139} (1978) 1.  
%
\bibitem{LMc}
M.~Lavelle and D.~McMullan, Phys. Lett. {\bf 329B} (1994) 68; {\it Constituent
Quarks From QCD}(Plymouth Preprint MS-95-06).

\bibitem{WC}
K. G. Wilson,  Phys. Rev. {\bf D10} (1974) 2445 . \\
M. Creutz, ``Quarks Gluons and Lattices,''  Cambridge University Press 1983. 


\end{thebibliography}
\end{document}